\documentclass[12pt,epsf,epsfig,psfig]{article}
\usepackage{graphicx}
\usepackage{epstopdf}
\usepackage{epsfig}
\usepackage{cite}
\def\C{c{\bar c}}

\def\b{b{\bar b}}
\def\q{q{\bar q}}

\def\g{\gamma_s}
\def\b{\bar \beta}

\def\be{\begin{equation}}
\def\ee{\end{equation}}

\def\lsim{\raise0.3ex\hbox{$<$\kern-0.75em\raise-1.1ex\hbox{$\sim$}}}
\def\gsim{\raise0.3ex\hbox{$>$\kern-0.75em\raise-1.1ex\hbox{$\sim$}}}

\def\PL{{ Phys.\ Lett.\ }}
\def\PR{{ Phys.\ Rev.\ }}

\def\ZP{{ Z.\ Phys.\ }}
\def\EP{{ Europ.\ Phys.\ J.\ C}}

\begin{document}


CERN-PH-TH/2013-251 \hfill 24.\ 10.\ 13.

~~\vskip0.7cm

\centerline{\Large \bf Causality Constraints on Hadron Production}

\bigskip

\centerline{\Large \bf In High Energy Collisions}

\vskip0.5cm 

\centerline{\bf Paolo Castorina$^{\rm a,b}$ and Helmut Satz$^{\rm c}$} 

\bigskip
\centerline{a: Dipartimento di Fisica ed Astronomia, Universita' di Catania, Italy}
\centerline{b: PH Department, TH Unit, CERN, CH-1211 Geneva 23, Switzerland}
\centerline{c: Fakult\"at f\"ur Physik, Universit\"at Bielefeld, Germany}



\vskip1cm

\centerline{\large \bf Abstract}

\bigskip

For hadron production in high energy collisions, causality requirements lead 
to the counterpart of the cosmological horizon problem: the production occurs 
in a number of causally disconnected regions of finite space-time size. As
a result, global\-ly conserved quantum numbers (charge, strangeness, baryon 
number) must be conserved locally in spatially restricted correlation 
clusters. This provides a theoretical basis for the observed suppression 
of strangeness production in elementary interactions ($pp$, $e^+e^-$). In 
contrast, the space-time superposition of many collisions in heavy ion 
interactions largely removes these causality constraints, resulting in 
an ideal hadronic resonance gas in full equilibrium.

\vskip1cm

\section{Introduction}

\begin{figure}[h]
\centerline{\psfig{file=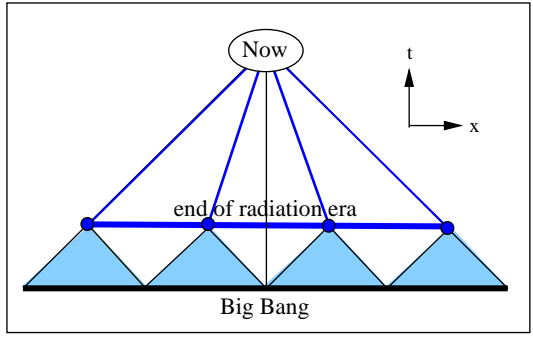,width=7cm}}
\caption{The origin of the observed cosmic background radiation}
\label{CBR}
\end{figure}

The temperature of the 
cosmic microwave background radiation (CBR) is, with a precision of
up to one part in $10^5$, found to be the same, 
some 2.7$~\!^{\circ}~\!$Kelvin, throughout the observable universe. 
This constitutes one of the basic problems of Hot Big Bang cosmology, since
at the end of the radiation era, when the CBR first appeared, the presently
visible universe consisted of a huge number of causally disconnected spatial 
regions; for a schematic view, see Fig.\ \ref{CBR}. How could such a 
uniformity in temperature arise without any communication between the 
radiation sources? 

\medskip 

The standard explanation has the equilibration arising either before or at 
inflation. In the inflation 
process, shortly after the Big Bang, the transition to the 
present stable vacuum ground state took place, accompanied by an exponential 
growth of the scale factor. This implies that when the present
constituents of matter and radiation first appeared in our world, they were
already in the same thermal state throughout all of space. They inherited 
this thermal behavior from a previous world of very much smaller dimension,
in which they were in causal contact and hence able to equilibrate. 

\medskip

\begin{figure}[h]
\centerline{\psfig{file=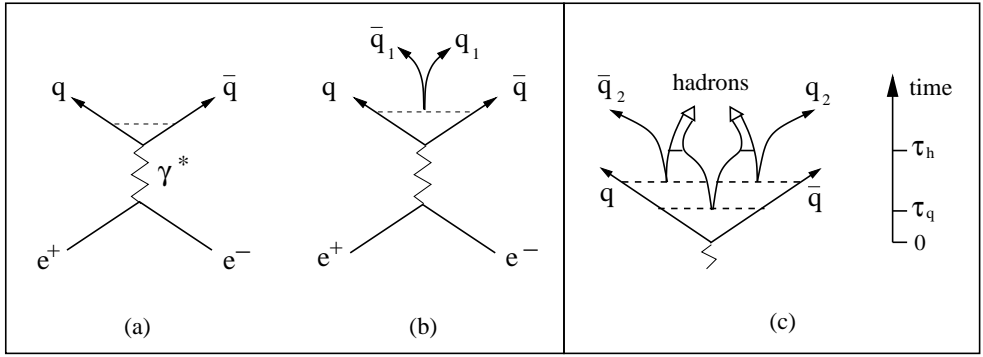,width=13cm}}
\caption{Initial stages of $e^+e^-$ annihilation}
\label{initial}
\end{figure}

The evolution of elementary high energy collisions is generally described 
in terms of an inside-outside cascade \cite{Bjorken}. It specifies a 
boost-invariant proper time $\tau_q$, at which local volume elements 
experience the transition from an initial state of frozen virtual partons
(``color glass'') to the on-shell partons which will eventually form 
hadrons. This partonisation time can be estimated most easily in
$e^+e^-$ annihilation (see Fig.\ \ref{initial}). 

\medskip

\begin{figure}[h]
\centerline{\psfig{file=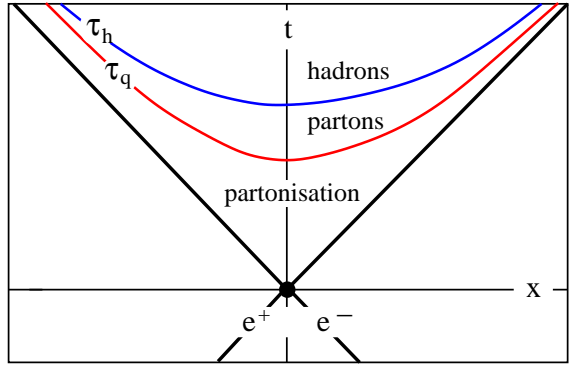,width=7.5cm}}
\caption{The evolution of $e^+e^-$ annihilation}
\label{parto-hadro}
\end{figure}

The initial quark-antiquark
pair is bound by a string of tension $\sigma$. When the separation
distance $x_q$ of the initial pair exceeds the energy $2 \omega_q$
of an additional $\q$ pair, the string breaks and the virtual pair is 
brought on-shell. For quarks of mass $m_q$, this energy is determined by
\be
\sigma x_q = 2\sqrt{m_q^2 + k_T^2}, 
\ee
where $k_T$ is the transverse momentum of each quark in the newly formed 
pair. Through uncertainty relations, this is given by $k_T=\sqrt{\pi \sigma
/2}$, leading to
\be
x_q \simeq \sqrt{2 \pi \over \sigma} \simeq 1~{\rm fm},
\ee
using $\sigma \simeq 0.2$ GeV$^2$ and $m_q \ll \sigma$. From this, 
we estimate
\be
\tau_q \simeq \sqrt{\sigma \over 2 \pi} \simeq 1~{\rm fm}.
\ee
This process is subsequently iterated, leading to a cascade of emitted
$\q$ pairs; while the first pair appears at rest in the center of mass
of the annihilation process, the subsequent pairs are produced at increasing 
rapidities. The different pairs will eventually bind to form free-streaming
hadrons; for a boost-invariant evolution, this defines a second time
threshold, the hadronisation time $\tau_h > \tau_q$. The overall scheme
is summarized in Fig.\ \ref{parto-hadro}. 

\medskip

\begin{figure}[h]
\centerline{\psfig{file=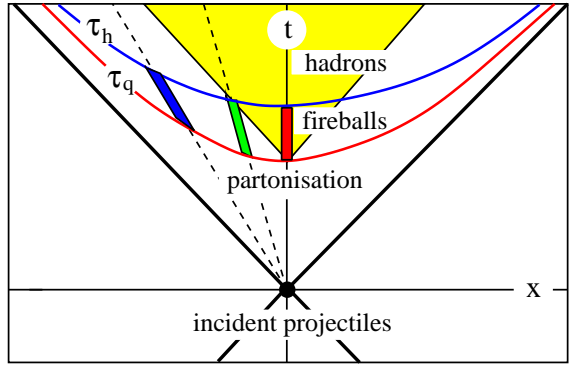,width=7.5cm}}
\caption{The boost-invariant evolution of a high energy collision, from 
the partonisation stage ($\tau \leq \tau_q$) to a fireball 
($\tau_q \leq \tau \leq \tau_h$) to hadrons ($\tau \geq \tau_h$). 
The region causally connected to a fireball at $\eta=0$ is colored in 
yellow, the fireball itself in red. An identical fireball for $\eta =\eta_d$ 
is marked in green, one for $\eta > \eta_d$ in blue.}
\label{causal}
\end{figure}

The generalization to $pp$ collisions is straight-forward: again there is 
a finite time $\tau_q$ needed to bring the partons on-shell, and after a 
larger time $\tau_h$, these combine to form hadrons. We denote the 
bubbles of medium for proper time $\tau$, with
$\tau_q < \tau <\tau_h$, as ``fireballs''.
Hadronisation thus occurs through the formation of partonic fireballs
in a cascade of increasing rapidities. 
In a boost-invariant scheme, the center of mass space-time coordinates $x,t$, 
with $x$ denoting the collision axis, are related to proper time $\tau$ 
and spatial rapidity $\eta$ through
\be
t=\tau \cosh \eta,~~~x= \tau \sinh \eta,
\label{2}
\ee
with $c=1$. The resulting evolution is illustrated in Fig.\ \ref{causal},
where the transition curves are determined by $t^2 - x^2 = \tau^2$. 
Schematically included in this figure is a fireball at $\eta=0$ and one 
at a larger $\eta$. 

\medskip

Both the partonisation and the hadronisation points 
for the system at larger $\eta$ are seen to be well outside the future region
of the $\eta=0$ fireball. More specifically, the hadronization point for
the large $\eta$ fireball just touches the event horizon of the $\eta=0$
fireball for
\be
\tanh \eta_d = { \tau_h^2 - \tau_q^2 \over \tau_h^2 + \tau_q^2},
\label{3}
\ee
defining the fireball range causally connected to the system at $\eta=0$. 
Beyond this rapiditiy, i.e., for $\eta > \eta_d$, the two fireballs 
are causally disconnected and cannot synchronize each other's thermal status.  
For the moment we are here neglecting the spatial extension of the
fireball, but we shall return to this aspect shortly.

\medskip

To illustrate, we choose $\tau_q=1$ fm and $\tau_h = 2$ fm; in this case,
a fireball with $\eta > 0.7 $ cannot communicate with one at $\eta=0$. The 
longer the fireball lifetime is, the larger is the rapidity range 
of fireballs in causal communication with that at $\eta=0$. The increase
of the range with fireball lifetime is quite slow, however; even for
$\tau_h = 7$ fm, the rapidity horizon is only $\eta_d=2$. In other words,
collisions at RHIC or at the LHC will lead to hadron production from 
causally disconnected fireballs.  

\medskip

The observation just made does not, of course, rule out a causal connection
(and hence correlations) for hadron production at large rapidity intervals; 
it only means that any correlations must have originated in the earlier 
partonisation stage. It does imply, however, that any state formed at
$\eta=0$ after a finite time interval cannot synchronize its thermal status
with a corresponding state at larger rapidity. We thus conclude that the 
fireballs 
formed in elementary high energy collisions appear in causally disconnected 
regions, which cannot communicate and thus in particular cannot establish a 
uniform temperature. If the hadronization temperature is found to be the 
same for different kinematic regions, this must be due to the local 
hadronization nature. There does not exist some large equivalent global 
system in thermal equilibrium, since any such equilibrium requires 
communication.

\section{Causal Connection of Fireballs}

\begin{figure}[h]
\centerline{\psfig{file=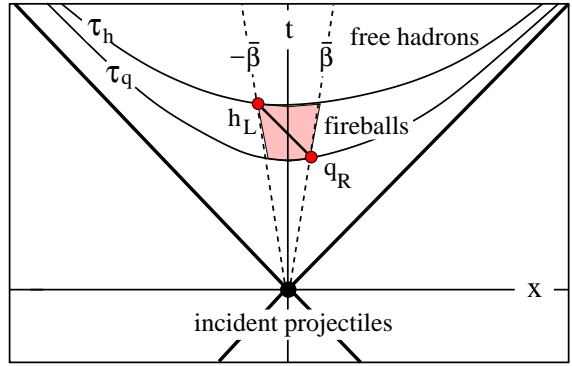,width=7.5cm}}
\caption{The formation and evolution of a fireball at rest in the
center of mass; the fireball evolution is indicated in pink.}
\label{evo}
\end{figure}

In the previous section, we had obtained in eq.\ \ref{3} the maximum 
rapidity $\eta_d$ for which a fireball could still receive a signal from a
one at $\eta=0$. Here the spatial extension of the fireball was for
simplicity neglected. For a more realistic situation, we have to consider a 
fireball of finite spatial extent. We take the longitudinal extension 
of the system to be
vanishingly small at the interaction time $t=0$; for sufficiently high
energy, this is expected to be a good approximation. The evolution of
the system is shown in Fig.\ \ref{evo}, where the shaded area defines
the fireball produced at rest in the CMS. The extremal velocity lines 
$\pm \beta = \pm v/c$ specify the spatial size of this fireball at the 
time $\tau_q$ of formation and its expansion up to the hadronisation 
time $\tau_h$. To consider
the system as one fireball, we require that the spatially right-most
point $q_R$ at formation can send a signal to the spatially left-most
point $h_L$ at hadronisation; i.e., we require that the most separate
points of the fireball can still communicate. This definition of a 
``causal'' fireball is evidently an upper limit in size; one may wish
to impose more stringent conditions and obtain a smaller fireball. 
We will keep that in mind in what follows. The crucial requirement in
our case is that the world-line connecting $q_R$ on the $\tau_q$ 
hyperbola with the point $h_L$ on the $\tau_h$ hyperbola is 
light-like, as shown in Fig.\ \ref{evo}.  

\medskip

To determine the resulting value of the velocity
$\beta$, we note that the point $q_R$ has the coordinates
\be
q_R = \left[ {\tau_q \over \sqrt{1-\beta^2}}, 
{\beta\tau_q \over \sqrt{1-\beta^2}} \right],
\label{4}
\ee
while $h_L$ is given by
\be
h_L = \left[ {\tau_h \over \sqrt{1-\beta^2}}, 
-{\beta\tau_h \over \sqrt{1-\beta^2}}\right].
\label{5}
\ee
The light ray eminating from $q_R$ is described by
\be
-\left(t -  {\tau_q \over \sqrt{1-\beta^2}}\right) =
x - {\beta \tau_q \over \sqrt{1-\beta^2}}.                 
\label{6}
\ee
Imposing that $h_L$ lies on this line leads to
\be
\beta = {\tau_h - \tau_q \over \tau_h + \tau_q}
\label{7}
\ee
and thus determines the rapidity with which the edges of the 
central fireball move out by its expansion. The resulting
spatial extension of this fireball becomes
\be
d =  {2 \beta \tau_h \over \sqrt{1-\beta^2}}
= \sqrt{{\tau_h \over \tau_q}}(\tau_h - \tau_q),
\label{8} 
\ee
measured at the time of hadronisation in the center of mass and thus in the 
proper frame of the fireball. This is the maximum initial size the fireball 
can have and still retain in its life-time a causal connection between its 
most distant space-time points. It is therefore fully determined by the
proper fireball formation time $\tau_q$ and its proper life-time 
$\tau_h - \tau_q$. In table \ref{size1}, we show the resulting velocities 
 $\beta$ and rapidities $\eta$  for the fireball edges and the
radii ($r=d/2$) of the fireballs produced at rest in the center of mass
at $\tau_q=1$ fm, for different values of the fireball life-time.
Of course the size of the fireball increases with increasing
hadronisation time; it is only the finite life-time of the partonic state 
that causes the total rapidity range for production to become divided into
causally disconnected segments. The rapidity extension of a fireball,
as we have defined it here, is roughly plus/minus one unit for 
$\tau_q=1$ fm, $\tau_h=3$ fm; its (proper) spatial radius at the time 
of formation is about 2 fm.
 
\medskip

\begin{table}
\begin{center}
\begin{tabular}{ccccc}
\hline \hline 
$\tau_h$ [fm]  &$\bar \beta$ & $\eta$ & $r$ [fm] & \\ 
\hline
$2~~~$& $ 0.33 ~~$ &$ 0.35 ~~$ & $ 0.7 $ 
\\
$3~~~$& $ 0.50 ~~$ &$ 0.55 ~~$ & $ 1.7 $ 
\\
$4~~~$& $ 0.60 ~~$ &$ 0.69 ~~$ & $ 3.0$
\\
$5~~~$& $ 0.67 ~~$ &$ 0.81~~$ & $ 4.5 $
\\
 \hline \hline
 \end{tabular}
\end{center}
${}$\vskip-.5cm \caption
{\label{size1} 
Velocity ($\beta$) and rapidity ($\eta$) limits of a fireball at 
rest in the center of mass, and its proper hadronisatin
radius $r$, as given
by eqs.\ \ref{7} and \ref{8}, for a formation time $\tau_q=1$ fm and
different hadronisation times $\tau_h$.}

\end{table}

\medskip

\begin{figure}[h]
\centerline{\psfig{file=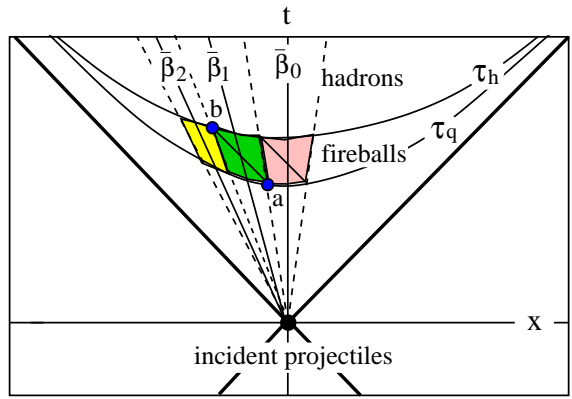,width=7.5cm}}
\caption{The formation and evolution of two fireballs, one at rest in the
center of mass (pink), one moving with velocity $\b_1$ (green). The light 
line from $a$ to $b$ determines the largest velocity difference still allowing 
a full causal connection between the two fireballs. For illustration, a third 
(yellow) fireball is shown, moving with velocity $\b_2$ and 
not causally connected to the one at rest in the center of mass.}
\label{evo2}
\end{figure}

We now assume complete boost invariance: the collision leads to the 
production of identical fireballs at all rapidities, with identical
formation and hadronisation times $\tau_q, \tau_h$ in their respective
rest frames. To study the causal connection of fireballs moving at
different rapidities, it is helpful to introduce a more specific notation
for their velocities. We denote the velocity of the fireball at rest in
the CMS by $\b_0=0$, and its extremal velocities by 
$\beta_{0L}=-\beta$ and 
$\beta_{0R}=\beta$. The neighboring fireball then has a central velocity
$\b_1$ and extremal velocities $\beta_{1L}$ and $\beta_{1R}$.
In its own rest-frame, this fireball
will have the same evolution pattern and spatial size as the one at
rest in the center of mass. To define a causal connection between this
fireball and the one at rest in the center of mass, we determine 
the largest value of $\b_1$, which still allows any point of the moving 
fireball to receive at (the latest) time $\tau_h$ a signal from at least
one point of the CMS fireball emitted at (the earliest) time $\tau_q$,
and vice versa, for the cms fireball. The relevant geometry is illustrated 
in Fig.\ \ref{evo2}. It is evident that the left extreme world-line
of the central fireball must then coincide with the right extreme of the 
fireball moving with velocity $-\b_1$. In other words, two adjacent fireballs
of identical structure will, in the sense just defined, be causally
connected. The next one ``down the line'', however, with velocity $\b_2$,
is causally disconnected from the central fireball. 

The determination of the central velocities of the successive fireballs is
given in the appendix; the result is
\be
\beta_n = {\tau_h^{2n} - \tau_q^{2n} \over \tau_h^{2n} + \tau_q^{2n}},
~~n=0,1,2,...
\label{9}
\ee 
Similarly, we obtain for the left extremal velocity of the n-th fireball
\be
\beta_{nL} = {\tau_h^{2n+1} - \tau_q^{2n+1} \over 
\tau_h^{2n+1} + \tau_q^{2n+1}},
~~n=0,1,2,...,
\label{9a}
\ee 
where $\beta_{0L}$ reduces to the value already given by eq.\ \ref{7}
for the fireball at rest in the overall center of mass.
Moreover, quite generally $\beta_{nL} = \beta_{(n+1)R}$. We have thus
divided the thermal space-time region, between $\tau_q$ and $\tau_h$,
into separate (non-overlapping) fireballs, such that next neighbors are 
causally connected, all further ones not.

\medskip

To illustrate the mesh of the net thus obtained, we list in table 
\ref{size2} the values of the velocities and rapidities of the first 
moving fireball, as measured in the CMS, for the fireball life-times 
used above. These values specify the maximum rapidity a moving fireball
can have and still remain causally connected to the one at rest in 
the CMS.

\begin{table}
\begin{center}
\begin{tabular}{ccccc}
\hline \hline 
$\tau_h$ [fm] &$\beta_1$ &$\eta_1$ \\ 
\hline
$2~~~$& $ 0.60 ~~$ &$ 0.70 ~~$  
\\
$3~~~$& $ 0.80 ~~$ &$ 1.10 ~~$  
\\
$4~~~$& $ 0.88 ~~$ &$ 1.39 ~~$ 
\\
$5~~~$& $ 0.92 ~~$ &$ 1.61~~$ 
\\
 \hline \hline
 \end{tabular}
\end{center}
${}$\vskip-.5cm 
\caption{Limiting velocities and rapidities for a moving fireball to
have causal connection with one at rest in the center of mass, see eq.(9)}
\label{size2}
\end{table}

\section{Hadronisation of Fireballs}

The hadrons formed through the final parton fusion constitute in principle
a complex interacting medium. A great simplification of this situation is
provided by an old argument \cite{BU,DMB}: if the interactions between the
basic hadrons, mesons and baryons, are resonance-dominated, then the 
interacting system of ground state hadrons can be replaced by an ideal
gas of all possible resonances. The relative abundances of the different
hadrons are in this case determined simply by the corresponding phase
space weights, specified in terms of the hadron masses and intrinsic
degrees of freedom.  

\medskip

The resulting statistical hadronisation model, based on a ideal gas of all 
observed hadronic resonances, provides an excellent general account for hadron 
production in high energy collisions, from $e^+e^-$ annihilation to the 
collision of heavy nuclei (see, e.g., \cite{Beca-Passa,PBM-R-St,Beca-LB,
BCMS}, and further references given there). All high energy data lead to 
a universal hadronisation temperature around 160 MeV, in accord with the
pseudo-critical temperature found in finite temperature lattice QCD 
with physical quark masses and for vanishing or low baryon density. This 
raises the question if and how hadronisation 
in elementary collisions differs from that in nucleus-nucleus interactions. 
Here the crucial observation is that in elementary collisions, the production 
of hadrons containing $n$ strange quarks or antiquarks is reduced in
comparison to the ideal resonance gas prediction. This reduction can be
accounted for by the introduction of a universal strangeness suppression 
factor $\g^n$, where $n$ denotes the number of strange quarks and/or
antiquarks contained in the hadron in question \cite{Rafelski}. The 
value of $\g$ is rather energy-independent 
and found to be around 0.5 to 0.7, from some 20 GeV up to LHC energies. 
In nuclear collisions, in contrast, $\g$ appears to converge to unity at 
RHIC and LHC energies, apart from slight corrections presumably due to 
corona effects \cite{corona1,corona2}. 

\medskip

The statistical hadronization model assumes that hadronization in high 
energy collisions is a universal process proceeding through the formation of 
multiple massive colorless clusters or fireballs of finite spacial 
extension and distributed over the rapidity range of the process.
These clusters are taken to decay into hadrons according to a purely 
statistical law: every multi-hadron state of the cluster phase space 
defined by its mass, volume and charges is equally probable. The mass 
distribution and the distribution of charges (electric, baryonic 
and strange) among the clusters and their (fluctuating) number are 
determined in the prior dynamical stage of the process. 
Hence in principle one would need the mentioned dynamical distributions 
in order to make definite quantitative predictions. However, for
Lorentz-invariant quantities such as multiplicities, 
one can further simplify matters by assuming
that the distribution of masses and charges among clusters is again
purely statistical, so that, as far as the calculation 
of multiplicities is concerned, the set of clusters becomes equivalent, 
on average, to one large cluster, the {\em equivalent global cluster},
whose volume is the sum of proper cluster volumes and whose charge is the 
sum of cluster charges, and thus the conserved charge of the initial 
colliding system. In such a global averaging process, the equivalent cluster 
in many cases turns out to be large enough in mass and volume so that the 
canonical ensemble becomes a good approximation. 

\medskip

To obtain a simple expression for our further discussion, we neglect
for the moment an aspect which is important in any actual analysis. 
Although in elementary collisions the conservation of the various discrete 
Abelian charges (electric charge, baryon number, strangeness, heavy flavour) 
has to be taken into account {\sl exactly} \cite{HR}, 
we here consider for the moment 
a grand-canonical picture. We also assume Boltzmann distributions for all 
hadrons. The multiplicity of a given scalar hadronic species $j$ then becomes

\be
\langle n_j \rangle^{\rm primary} = \frac{V T m_j^2}{2\pi^2} 
\g^{n_j} {\rm K}_2\left(\frac{m_j}{T}\right)\, 
\ee

with $m_j$ denoting its mass and $n_s$ the number of strange quarks/antiquarks
it contains. Here primary indicates that it gives the number at the
hadronisation point, prior to all subsequent resonance decay.
The Hankel function $K_2(x)$, with $K(x) \sim 
\exp\{-x\}$ for large $x$, gives the Boltzmann 
factor, while $V$ denotes the overall equivalent cluster volume. In other 
words, in an analysis of $4 \pi$ data of elementary collisions, $V$ is the 
sum of the all cluster volumes at all different rapidities. It thus scales
with the overall multiplicity and hence increases with collision energy.  
A fit of production data based on the statistical hadronisation model thus
involves three parameters: the hadronisation temperature $T$, the 
strangeness suppression factor $\gamma_s$, and the equivalent global cluster
volume $V$. 

\medskip

We want to use the results of 
the present paper to show that the nature of $V$ in elementary
collisions is quite different from that in nuclear collisions, and this 
can in effect lead to different behavior in the two cases. Strangeness
production is perhaps the most readily accessible such phenomenon. In
elementary collisions, the clusters at rapidities sufficiently far apart 
are, as we have seen, causally disconnected, so that they cannot exchange
information. Hence strangeness must be conserved locally; in $pp$
collisions, for example, each cluster must have strangeness zero.
Thus typically there will be only one pair of strange particles within a 
given cluster, adding up to zero total cluster strangeness.
Such a local strangeness conservation is known to lead to a 
suppression of strangeness production \cite{HRT}; we return to 
details shortly.
In high energy nuclear collisions, on the other hand, the equivalent global 
cluster consists of the different clusters from the different nucleon-nucleon 
interactions at a common rapidity. At mid-rapditiy, for example, we thus 
have the sum of the 
superimposed mid-rapidity clusters from the different nucleon-nucleon 
collisions, and these are all causally connected, allowing strangeness 
exchange and conservation between the different clusters.    

\medskip

As noted, the local conservation of charges, and in particular of strangeness, 
has in fact been proposed for quite some time as the mechanism responsible for 
strangeness suppression \cite{HRT}; more details are given in appendix A2.
In the grand canonical
approach, the introduction of the suppression factor $\gamma_s$ achieves
the observed suppression. The alternative of local strangeness suppression
is based on two features. First, one imposes exact strangeness conservation,
which leads to a volume-dependent strangeness reduction 
\cite{HR,BRSt}; the 
ratio of canonical to grand-canonical partition functions,
\be
{Z_{\rm can}(T,V,S) \over Z_{\rm gcan}(T,V,\langle S \rangle)} < 1
\label{can}
\ee
approaches unity only in the limit of large volumes. However, 
in elementary collisions with the corresponding
overall equivalent cluster volume, the resulting reduction is not sufficient 
to account for the observed strange particle rates. Hence it was argued that
if in a given collision only one pair of strange hadrons is produced, 
these should appear close to each other spatially, the more so if
the medium is relatively short-lived. This approach thus introduces
somewhat {\sl ad hoc} a strangeness correlation volume $V_c<V$, within 
which strangeness has to be conserved exactly. The corresponding model 
thus now has $T$, $V$ and $V_c$ as the parameters to be specified by 
the data, and fits based on such a model provide as good an account for 
the data as the earlier $\gamma_s$ scheme \cite{kraus1,kraus2}, with
the exception of the $\phi$, to which we return later. 
However, {\sl a priori} little is known about $V_c$, and in particular 
it remains open what happens to it in nuclear collisions. 

\medskip

We here propose that the strangeness correlation volume $V_c$ is in fact
that of a causally connected cluster; causal connectivity thus provides the 
fundamental reason for local strangeness conservation and hence for the 
strangeness suppression observed in elementary interactions.
It is moreover clear that in nucleus-nucleus interactions, the overlapping
fireballs produced at fixed rapidity by the different nucleon-nucleon
collisions will give rise to a much larger causally connected volume and
thus effectively remove the locality constraints. Moreover, if {\sl very} high
energy $pp$ interactions lead to multiple jet production, this could
eventually lead to a similar effect, with overlapping clusters from the
different jet directions. 

\medskip

We have seen how the size of the causally connected cluster volumes 
varies with the fireball life-time. An obvious question therefore is
whether the fits to production data lead to reasonable cluster sizes.
It is found \cite{kraus1,kraus2} that good fits to data at $\sqrt s=17.3$ 
and 200 GeV require a strangeness correlation radius of about 1 fm, 
while leading to the same universal hadronisation temperature of about 
160 MeV. In our considerations, this is seen to be in accord with a 
hadronisation time $\tau_h$ of about 2 - 3 fm. For an evolution of 
the kind shown in Fig.\ \ref{initial} that makes good sense: it takes 
about 1 fm to form the first $\q$ pair, and another to have it hadronize. 
The causality constraints in elementary high energy collisions thus 
appear to provide the reason for the observed strangeness suppression, 
thereby justifying the strangeness correlation model \cite{HRT}. 

\medskip

We further note here that in elementary collisions, hadronisation as Unruh 
radiation arising from quarks tunnelling through their color confinement 
horizon \cite{CKS} inherently contains locals strangeness conservation. In 
such a scheme, the maximum separation between $s$ and $\bar s$ can never 
exceed the hadronic scale leading to string breaking, i.e., about 1 fm,
thus enforcing strangeness production in a very restricted spatial volume. 

\medskip

In the case of heavy nuclei, on the other hand, we find in the center 
of mass with increasing collision energy a superposition of more and 
more individual nucleon-nucleon interactions in the same space-time 
region. At high enough collision
energy, there will thus be on the average around five or six superimposed 
nucleon-nucleon collisions, so that there now exists a causally connected 
region having an effective volume five or six times larger than that in a 
nucleon-nucleon collision, with a corresponding increase in the number of 
produced strange particles. An $s$ quark produced in any specific 
nucleon-nucleon collision now finds so many $\bar s$ from other such 
collisions in its immediated environment that no spatial constraints on its 
partner $\bar s$ are necessary. Moreover, the superposition of collisions
at central rapidity greatly increases the partonic density there. As a
consequence, it takes a longer time for the system to expand up to the
hadronisation point, so that $\tau_h$ now is considerably larger. This
aspect further increases the correlation volume. In terms of a conventional
statistical description, it implies that $\gamma_s$ is driven towards unity. 
At $\sqrt s = 17.3$ GeV, the overlap is not yet complete: when the first 
nucleons collide, those at the opposite edges of the two nuclei are still 
some 3 fm apart, so that we can still expect some strangeness suppression,
and this is indeed observed. At $\sqrt s = 200$ GeV, this separation
has decreased to 0.3 fm, so that at RHIC and at the LHC, there should
not be any suppression, apart from possible corona contributions.

\section{The Problem of Hidden Strangeness}

The approach presented here for the suppression of strangeness production
in elementary collisions contains one open issue, which arises in all
attempts of local strangeness conservation. It is found experimentally that
the $\phi$ meson, consisting of an $s \bar s$ pair, is also suppressed,
although from a hadronic point of view, it is of zero strangeness. In the 
conventional statistical model with a strangeness suppression factor,
the power of $\gamma_s$ is determined by the number of $s$ plus $\bar s$
quarks a given hadron contains. Hence the $\phi$ gets a factor $\gamma_s^2$, 
which leads to rough agreement with the data. In contrast, in a canonical 
formulation on a hadronic level, the $\phi$ does not present any quantum 
number to be conserved exactly and is not subject to any suppression.
There are (at least) two ways to resolve this issue.

\medskip

It is well-known that
quarkonia ($\C$ and $b \bar b$ mesons, states of hidden heavy flavor) cannot 
be accommodated at all in any statistical approach. Their production and
binding is governed by gluon dynamics instead. One might therefore argue 
that the $\phi$, as hidden strangeness meson, also falls into this
category and hence its abundance cannot be determined in a statistical  
model. However, such an approach has its own problems. The abundance
of charmonium and bottomonium states is {\sl underpredicted} by orders of
magnitude, while that of the $\phi$ is {\sl overpredicted} by a factor
four or so. The quarkonium states are below the open charm/beauty thresholds,
while the $\phi$ decays strongly into a $K \bar K$ pair. Finding a common
ground for it and the quarkonia is therefore surely not easy.

\medskip

Another approach is that taken in the introduction of the suppression
factor $\gamma_s$ in powers of the content of strange {\sl and} antistrange
quarks. The evolution of the statistical hadronisation went from 
grand-canonical to canonical, and on to the introduction of a correlation
volume in the hadronic canonical formulation. The disagreement of the
$\phi$ abundance may thus be nature's way of telling us that strangeness
correlation really occurs already on a pre-hadronic level. Requiring
exact strangeness conservation for the quark system in the fireball prior 
to hadronisation would in fact result in canonical strangeness
suppression of both open and hidden strangeness (see Fig.\ \ref{casto6}), 
of a functional form very similar to that obtained on a hadronic level 
in Appendix 2.

\bigskip

\begin{figure}[h]
\centerline{\psfig{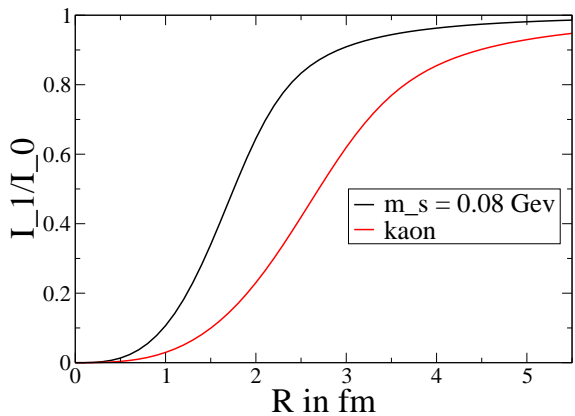}}
\caption{The suppression factor for exact strangeness conservation
of strange quarks of mass $m_s=80$ MeV in a volume of radius $R$,
compared to the suppression factor of kaons (see appendix 2), both 
at a temperature of 160 MeV.}
\label{casto6}
\end{figure}
\vskip1cm

{\bf Acknowledgements} The authors thank F.Becattini and U.Wiedemann for stimulating discussions. P.C. thanks the CERN TH-unit for the hospitality.

\bigskip

{\Large \bf Appendix A1}

\bigskip

According to the criterium of causal connection discussed in Section 2, 
the world-lines of the left extremum and of the right extremum of the 
fireball $n$, with $n \ge 1$, in the region $x<0$ are, respectively, 
$-\beta_{n+1} t=x$ and $-\beta_n t=x$ (see Fig.\ \ref{evo2}).

\medskip

The right extremum meets the hyperbola of the plasma formation time, 
$t^2-x^2 =\tau_q^2$, at the event point $E_a$ with coordinates
\be
(t^q_n, x^q_n) = (\frac{\tau_q}{\sqrt{1- \beta^2_n}} ,  
\frac{-\beta_n\tau_q}{\sqrt{1- \beta^2_n}})\label{a1}.
\ee
The light ray originating from the point $E_a$ has equation
\be
-t+\frac{\tau_q}{\sqrt{1- \beta^2_n}} = x - \frac{\beta_n\tau_q}
{\sqrt{1- \beta^2_n}}\label{a2}
\ee
and crosses the hyperbola of the hadronization time, $t^2-x^2 = \tau_h^2$, 
at the event point $E_b$ with coordinates
\be
(t^h,x^h) = \left({1\over 2 \tau_q}\sqrt{1+\beta_n \over 1 -\beta_n}
\left[\tau_h^2 + \tau_q^2 \frac{1-\beta_n}{1+\beta_n}\right], 
{1\over 2 \tau_q}\sqrt{1+\beta_n \over 1 -\beta_n}
\left[\tau_h^2 - \tau_q^2 \frac{1-\beta_n}{1+\beta_n}\right]\right).
\label{a3}
\ee
Since the event $E_b$ has to be on the world line of the left extremum, 
we must have $\beta_{n+1} = - x^h/t^h$, i.e.
\be
\beta_{n+1} = \left(\tau_h^2 - \tau_q^2{1-\beta_n \over 1+\beta_n}\right)
{\bigg/} 
\left(\tau_h^2 + \tau_q^2 {1-\beta_n \over 1+\beta_n}
\right),
\label{a4}
\ee
expressing our condition of causal connection. The first extremal world-line 
in the $x<0$ region has speed (see eq.(\ref{7}))
\be
\beta_1=\frac{\tau_h - \tau_q}{\tau_h + \tau_q},\label{a5};
\ee
hence $(1-\beta_1)/(1+\beta_1)= \tau_q/\tau_h$, and, by iteraction, the 
speeds of the left extrema are found to be $(n \ge 0)$
\be
\beta_{n+1} = \frac{\tau_h^{2n+1} - \tau_q^{2n+1}}{\tau_h^{2n+1} 
+ \tau_q^{2m+1}}\label{a6}
\ee
The speed of the cms of the $n$-th fireball, $\bar \beta_n$, with respect 
to the rest frame (corresponding to the cms of the fireball at $x=0$) 
is defined 
by requiring that all fireballs have the same structure in their cms. In 
other words, if $\beta_{n+1}$ and $\beta_n$ are the speeds of the two extrema 
of a fireball $n$ in the overall rest frame, and  $\beta_{n+1}^{'}$ and 
$\beta_n^{'}$ are the corresponding velocities in the rest frame of
this fireball, the speed $\bar \beta_n$ of the cms of the  
fireball with respect to the overall rest frame must be such that  
$\beta_{n+1}^{'} = - \beta_n^{'}$.
By the velocity composition law, it turns out that
\be
\bar \beta_n = \frac{(1+\beta_{n+1} \beta_n) - 
\sqrt{(1-\beta_{n+1}^2)(1-\beta_n^2)}}{\beta_{n+1} + \beta_n}\label{a7}
\ee
By eq.\ (\ref{a4}) and after some algebra, one obtains
\be
\bar \beta_n = {\left[\tau_h(1+\beta_n) - \tau_q (1-\beta_n)\right]^2 
\over \tau_h^2(1+\beta_n)^2  - \tau_q^2(1-\beta_n)^2}
\label{a8}.
\ee
By use of eq.\ (\ref{a5}) and some iteration, this gives
\be
\bar \beta_{n} = \frac{\tau_h^{2n} - \tau_q^{2n}}{\tau_h^{2n} 
+ \tau_q^{2n}}\label{a9}.
\ee
for the speed of fireball $n$.

\vskip1cm

{\Large \bf Appendix A2}

\bigskip

We here want to illustrate in some detail the mechanism of local
strangeness reduction. To simplify matters, let us assume that there
are only two hadron species: scalar and electrically neutral mesons, ``pions'' 
of mass $m_{\pi}$, ``kaons'' of mass $m_K$ and strangeness $s= 1$ together
with ``antikaons'' of the same mass but strangeness $s=-1$.
In this case, the grand canonical partition function for a system of 
of volume $V$ and temperature $T$ has the form
\be
Z_{GC}(T,V,\mu) = {VT \over 2\pi^2} \left[m_{\pi}^2 K_2(m_{\pi}/T)
+ m_K^2 K_2(m_K/T) e^{\mu /T}
+ m_K^2 K_2(m_K/T) e^{-\mu /T}
\right],
\label{A1}
\ee
where $\mu$ denotes the chemical potential for strangeness. If the
overall strangeness is zero, $\mu=0$ and the average density of mesons 
of type $i$ ($i=\pi,K,\bar K$) is given by
\be
n_i(T) = {T m_i^2 \over 2 \pi^2} K_2(m_i/T),
\label{A2}
\ee
while the ratio of kaon to pion multiplicities becomes
\be
{N_K \over N_{\pi}} = \left({m_K \over m_{\pi}}\right)^2
{K_2(m_K/T) \over K_2(m_{\pi}/T)} \simeq 
\left({m_K \over m_{\pi}}\right)^{3/2} \exp\{-{(m_K - m_{\pi})\over T}\}.
\label{A3}
\ee 
Both species densities and ratios thus are independent of the overall volume
$V$; they are determined by the respective masses and the hadronisation
temperature $T$. The grand canonical form assures that the average overall
strangeness is zero, but only the {\sl average}; there are fluctuations,
and, for example, the second cumulant
\be
\left({\partial^2 \ln~\!Z_{GC} \over \partial \mu^2}\right) \sim 
\langle S^2 \rangle
\label{A4} 
\ee
indicates that the average of the squared strangeness does not vanish.

\medskip

The grand canonical ensemble effectively corresponds to an average over
all possible strangeness configurations, with $\exp({\pm}\mu/T)$ as
weights. If instead we insist that the overall strangeness is {\sl exactly}
zero, we have to project out that term of the sum. This canonical ensemble
can lead to a
severe restriction of the available phase space and hence of the production 
rate. Thus the canonical density of kaons becomes \cite{HRT,BRSt}
\be
\tilde n_K(T,V) = n_K(T) {I_1(x_K) \over I_0(x_K)},
\label{A5}
\ee
where $I_n(X)$ is the $n-th$ order Bessel function of imaginary argument and
$n_K(T)$ is given by eq.\ (\ref{A2}) and 
\be
x_K = {VTm_K^2 \over 2 \pi^2} K_2(m_K/T).
\label{A6}
\ee

\medskip

\begin{figure}[h]
\centerline{\psfig{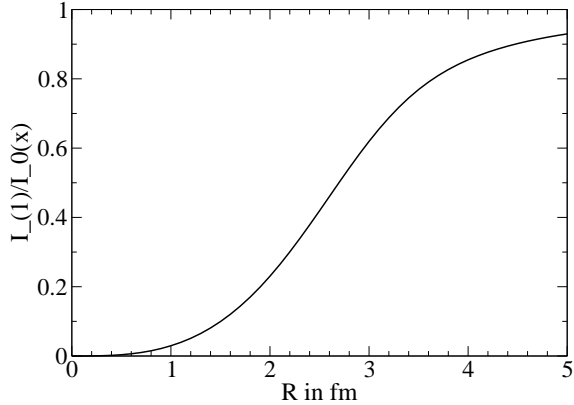}}
\caption{The suppression factor for exact strangeness conservation
of kaons in a volume of radius $R$.}
\label{casto2}
\end{figure}

\medskip

The canonical density, in contrast to the grand canonical form, thus
depends on the volume $V$ of the system.
Since $I_n(x) \sim x^n$ for $x \to 0$ and $I_n(x) \to e^x$ for $x \to \infty$,
we see immediately that in the large volume limit,
\be
\tilde n_K(T,V) \to n_K(T),
\label{A7}
\ee
the canonical form converges to the grand canonical one, as expected.
In the small volume limit, however, the Bessel function ratio results
in a strong suppression of canonical relative to grand canonical form,
with $I_1(x)/I_0(x) \to 0$ for $x\to 0$. For the actual values of the kaon
mass and $T \simeq 160$ MeV, this suppression sets in for volumes of
radii less than some 2 - 3 fm; above that, the grand canonical form
becomes valid. The form of the suppression factor is shown in Fig.\
\ref{casto2}; we recall that strange baryons are neglected in
obtaininng eq.\ \ref{A5}, so that the figure is for illustration only.

\medskip

We thus see that the exact conservation of charges, such as strangeness,
results in a ``canonical suppression'' for sufficiently small volumes.
Now the overall volume $V$ in the conventional description of $e^+e^-$ 
annihilation or in $pp$ collisions is that of the equivalent cluster
and hence determined largely by the number of pions. Thus imposing 
exact strangeness conservation here is not the solution - the total
volume is so large that there is no effective canonical suppression.
To obtain the observed strangeness reduction, an additional mechanism is
necessary. 

\medskip

This was obtained \cite{HRT}
by arguing that in the case of very few charge carriers,
charge neutralisation must occur in a correlation volume $V_c$ very much
smaller than the overall volume $V$. For a given charge, there must be 
an opposite charge nearby, not some large distance away. This argument
was supported by kinetic studies, indicating that the typical life-time
of the partonic medium is not sufficient for far-away charges to meet,
making exact conservation unlikely. As a result, the partition function 
for our pion-kaon system now becomes for exact strangeness zero
\be
Z(T,V,V_c) = {VT \over 2\pi^2} \left[m_{\pi}^2 K_2(m_{\pi}/T)
+ 2 m_K^2 K_2(m_K/T) 
\left({I_1(x_K(T,m_K,V_c)) \over I_0(x_K(T,m_K,V_c))}\right)\right],
\label{A8}
\ee 
where the argument of the Bessel functions is given by
\be 
x_K = {V_c T m_K^2 \over 2 \pi^2} K_2(m_K/T).
\label{A9}
\ee
and thus contains the strangeness correlation volume $V_c$ as further
parameter. By tuning $V_c$, we can thus achieve as much strangeness
suppression as desired.

\medskip

As mentioned, our considerations here are only meant as illustration.
In actual studies, both normal, strange and multi-strange baryons
have to be included, as well as all higher resonant states. If this
is done, a formulation of the type discussed here leads with a correlation
radius $R_c$ around 1 fm to the observed suppression and to a model which
can account for the data from elementary collisions as well as the
conventional $\gamma_s$ approach, except for the mentioned 
$\phi$ problem.

\end{document}